%% file: main.tex
\documentclass[sigconf]{acmart}
\acmSubmissionID{52}
\input{Preamble/packages}
\input{Preamble/definitions}

\input{Preamble/copyright}
\input{Preamble/authors}
\begin{document}
\title{Distilling a Small Utility-Based Passage Selector to Enhance Retrieval-Augmented Generation}

\begin{abstract}
Retrieval-augmented generation (RAG) enhances large language models (LLMs) by incorporating retrieved information. Standard retrieval process prioritized relevance, focusing on topical alignment between queries and passages. In contrast, in RAG, the emphasis has shifted to utility, which considers the usefulness of passages for generating accurate answers.
Despite empirical evidence showing the benefits of utility-based retrieval in RAG, the high computational cost of using LLMs for utility judgments limits the number of passages evaluated. 
This restriction is problematic for complex queries requiring extensive information. To address this, we propose a method to distill the utility judgment capabilities of LLMs into smaller, more efficient models. Our approach focuses on utility-based selection rather than ranking, enabling dynamic passage selection tailored to specific queries without the need for fixed thresholds. We train student models to learn pseudo-answer generation and utility judgments from teacher LLMs, using a sliding window method that dynamically selects useful passages. Our experiments demonstrate that utility-based selection provides a flexible and cost-effective solution for RAG, significantly reducing computational costs while improving answer quality. We present the distillation results using Qwen3-32B as the teacher model for both relevance ranking and utility-based selection, distilled into \rank$_{1.7B}$ and \utility$_{1.7B}$. 
Our findings indicate that for complex questions, utility-based selection is more effective than relevance ranking in enhancing answer generation performance. We will release the relevance ranking and utility-based selection annotations for the MS MARCO dataset, supporting further research in this area.
Our code and datasets can be found at \url{https://github.com/Trustworthy-Information-Access/UtilitySelection}.

\end{abstract}

\maketitle
% \authornotetext[1]{Corresponding authors.}
% \renewcommand{\thefootnote}{\fnsymbol{footnote}}
% \footnotetext[1]{Corresponding authors}
% \renewcommand{\thefootnote}{\arabic{footnote}}
% \setcounter{footnote}{0}
% \footnotetext{$^{*}$Corresponding authors}
\input{Sections/Introduction}

\input{Sections/Related_work}

\input{Sections/Method}

\input{Sections/Experiments}
\input{Sections/Futher_analyse}

\input{Sections/Conclusion}
\begin{acks}
This work was funded by the National Natural Science Foundation of China (NSFC) under Grant No. 62302486, the Innovation Funding of ICT CAS under Grant No. E361140, the CAS Special Research Assistant Funding Project, the project under Grants No. JCKY2022130C039, the Strategic Priority Research Program of the CAS under Grants No. XDB0680102, and the NSFC Grant No. 62441229.
\end{acks}
\bibliographystyle{ACM-Reference-Format}
\balance
\bibliography{sample-base}
% \bibliography
\end{document}

%% file: Preamble/packages.tex
\usepackage{color}
\usepackage{bbm}
\usepackage{multirow}
\usepackage[inline]{enumitem}
\usepackage{subfigure} 
\usepackage{sistyle}
\usepackage[ruled]{algorithm2e}
\usepackage{booktabs}
\usepackage{caption}
\SIthousandsep{,}
\usepackage{makecell}
\usepackage[skip=0pt]{caption}
\usepackage{amsmath}
\usepackage{hyperref}
\usepackage{array} 
\usepackage{graphicx}
\usepackage{xcolor}
\usepackage{acronym}
\usepackage[export]{adjustbox}
\usepackage{balance}

%% file: Preamble/definitions.tex
\newcommand{\heading}[1]{\vspace*{0.5mm}\noindent\textbf{#1.}}

\AtBeginDocument{%
  \providecommand\BibTeX{{%
    \normalfont B\kern-0.5em{\scshape i\kern-0.25em b}\kern-0.8em\TeX}}}

\makeatletter
\g@addto@macro\normalsize{%
  \abovedisplayskip 2pt plus1pt %minus1pt%
  \belowdisplayskip 2pt plus1pt
  \abovedisplayshortskip  2pt plus1pt%
  \belowdisplayshortskip  1pt plus1pt% minus1pt%
}
\makeatother

\newcommand{\rank}{RankQwen}
\newcommand{\utility}{UtilityQwen}

\acrodef{IR}{information retrieval}
\acrodef{LLM}{large language model}

%% file: Preamble/copyright.tex
\AtBeginDocument{%
  \providecommand\BibTeX{{%
    Bib\TeX}}}

\copyrightyear{2025}
\acmYear{2025}
\setcopyright{acmlicensed}
\acmConference[SIGIR-AP '25] {Proceedings of the 2025 Annual International ACM SIGIR Conference on Research and Development in Information Retrieval in the Asia Pacific Region}{December 7--10, 2025}{Xi'an, China.}
\acmBooktitle{Proceedings of the 2025 Annual International ACM SIGIR Conference on Research and Development in Information Retrieval in the Asia Pacific Region (SIGIR-AP '25), December 7--10, 2025, Xi'an, China}
\acmISBN{979-8-4007-2218-9/2025/12.}
\acmDOI{10.1145/3767695.3769496}
\ccsdesc[500]{Information systems~Question answering}
\ccsdesc[500]{Information systems~Novelty in information retrieval}
\ccsdesc[500]{Information systems~Top-k retrieval in databases}

\keywords{Relevance Ranking, Utility-based Selection, RAG}
\received{20 February 2007}
\received[revised]{12 March 2009}
\received[accepted]{5 June 2009}

%% file: Preamble/authors.tex
\author{Hengran Zhang}
\orcid{0009-0004-1144-1298}
\affiliation{
    \institution{State Key Laboratory of AI Safety, Institute of Computing Technology, Chinese Academy of Sciences} 
	\institution{University of Chinese Academy of Sciences}
	\city{Beijing}
	\country{China}
}
\email{zhanghengran22z@ict.ac.cn}

\author{Keping Bi}
\authornote{Keping Bi and Jiafeng Guo are the corresponding authors.}
\affiliation{
    \institution{State Key Laboratory of AI Safety, ICT, Chinese Academy of Sciences} 
	\institution{University of Chinese Academy of Sciences}
	\city{Beijing}
	\country{China}
}
\email{bikeping@ict.ac.cn}

\author{Jiafeng Guo}
\orcid{0000-0002-9509-8674}
\authornotemark[1]
\affiliation{
    \institution{State Key Laboratory of AI Safety, Institute of Computing Technology, Chinese Academy of Sciences} 
	\institution{University of Chinese Academy of Sciences}
	\city{Beijing}
	\country{China}
}
\email{guojiafeng@ict.ac.cn}

\author{Jiaming Zhang}
\affiliation{
	\institution{Baidu Inc}
 \city{Beijing}
 \country{China}
}
\email{zhangjiaming04@baidu.com}

\author{Shuaiqiang Wang}
\affiliation{
	\institution{Baidu Inc}
 \city{Beijing}
 \country{China}
}
\email{wangshuaiqiang@baidu.com}

\author{Dawei Yin}
\affiliation{
	\institution{Baidu Inc}
 \city{Beijing}
 \country{China}
}
\email{yindawei@acm.org}

\author{Xueqi Cheng}
\orcid{0000-0002-5201-8195}
\affiliation{
    \institution{State Key Laboratory of AI Safety, Institute of Computing Technology, Chinese Academy of Sciences} 
	\institution{University of Chinese Academy of Sciences}
	\city{Beijing}
	\country{China}
}
\email{cxq@ict.ac.cn}

\renewcommand{\shortauthors}{Hengran Zhang et al.}

%% file: Sections/Introduction.tex
\section{Introduction}

% \begin{figure}[h]
%   \centering
%   \includegraphics[width=\linewidth]{images/introduction.pdf}
%   \caption{Different answer generation performance (generator: Llama3.1-8B) directly with different top-$k$ retrieval results (retriever: BM25).}
%   \label{fig:original_ranked_answer}
% \end{figure}
\begin{figure}[h]
  \centering
  \includegraphics[width=\linewidth]{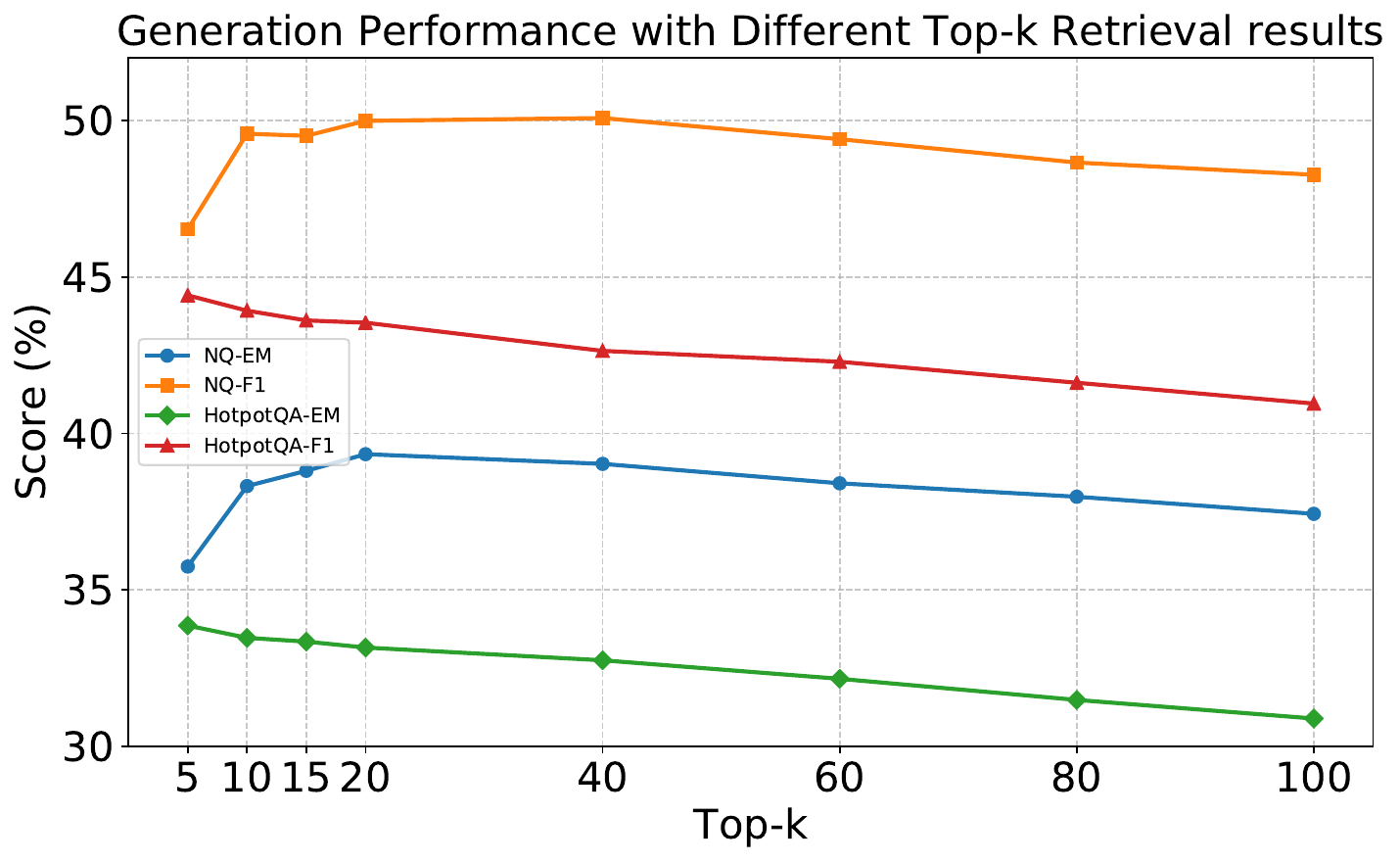}
  \caption{Different answer generation performance (generator: Llama-3.1-8B-Instruct) directly with different top-$k$ retrieval results (retriever: BM25).}
  \label{fig:original_ranked_answer}
\end{figure}
Retrieval-augmented generation (RAG) leverages retrieved information as external knowledge to empower large language models (LLMs) to answer questions. 
The criterion for measuring whether a result is helpful for RAG has shifted from relevance to utility \cite{shi2023replug,ke2024bridging,zhang2024large,zhang2024iterative}. 
Relevance typically focuses on the topical matching between a query and retrieved passages \cite{saracevic1988study, schamber1988relevance}. 
Utility, in contrast, emphasizes the usefulness of a passage in facilitating the generation of an accurate and comprehensive answer to the question \cite{zhang2025leveraging}. 
% Studies \cite{zhang2024large, zhang2024iterative} have also established this distinction, and 
Empirical results demonstrate that using retrieval results judged as having utility by LLMs in RAG can enhance the quality of subsequent answer generation \cite{zhang2024large,zhang2024iterative}.

Due to the high computation cost, using LLMs for utility judgments usually takes 10 to 20 passages as context \cite{zhang2024large,zhang2024iterative}. This is insufficient for weaker retrievers that rank useful passages at lower positions, and when complex questions require many useful documents to generate comprehensive answers. 
Although it is promising to scale utility judgments to a large number of candidate passages for RAG, using LLMs to do so is cost-prohibitive. 
% Current studies \cite{zhang2024large, zhang2024iterative} typically restrict utility judgments to approximately 20 passages per query due to these constraints. 
% This limitation is particularly problematic for complex queries, such as multi-hop reasoning questions, which often require synthesizing information from numerous passages to generate a complete answer. A limit of around 20 passages may be insufficient for such cases. 
Therefore, we propose to distill the utility judgment capability of LLMs to smaller models that are efficient to do so.   
% a critical research direction is transferring the capability for utility judgment into smaller, more efficient models. 

In this paper, we focus on utility-based selection rather than ranking when distilling smaller models. There are two reasons: 1) \citet{ke2024bridging} find that for effective RAG, the ranking of input passages is less critical than effectively filtering out low-quality passages. 2) 
% While utility itself can be used to rank passages, 
The number of passages that should be selected for different questions can vary. As shown in Figure \ref{fig:original_ranked_answer}, the optimal number of passages used for simple questions (i.e., in NQ) and complex questions (i.e., in HotpotQA) is different. 
If we conduct utility ranking, 
% it needs to determine the number of passages to use for answer generation. 
a fixed threshold is usually used, which introduces a hyperparameter to tune and can be suboptimal. 
% inherently static and fails to adapt to the specific demands of individual queries. 
% As shown in Figure \ref{fig:original_ranked_answer}, the optimal passage number is different for different queries. 
In contrast, utility-based selection mechanisms can dynamically determine how many passages to retain. 
Consequently, our goal is to distill a small utility-based selector from LLMs that are competent in zero-shot utility-based selection. 
% our core motivation is to distill the capability for utility-based selection into smaller, more efficient models, enabling scalable filtering for real-world deployment. 

% Large Language Models (
There have been several studies on distilling the zero-shot listwise ranking capability of LLMs (e.g., ChatGPT, GPT-4) into smaller efficient rankers \cite{sun2023chatgpt,pradeep2023rankvicuna, pradeep2023rankzephyr, liu2025leveraging, reddy2024first}, such as RankVicuna \cite{pradeep2023rankvicuna} and RankMistral \cite{liu2025leveraging}. 
% LLMs demonstrate significant potential for advancing relevance ranking, particularly through zero-shot listwise re-ranking techniques \cite{sun2023chatgpt}. 
% Consequently, numerous studies \cite{pradeep2023rankvicuna, pradeep2023rankzephyr, liu2025leveraging, reddy2024first} have successfully distilled the ranking capabilities of powerful LLMs (e.g., ChatGPT, GPT-4) into smaller, efficient models such as RankVicuna \cite{pradeep2023rankvicuna} and RankMistral \cite{liu2025leveraging}. 
The distilled student rankers ingest a large number of retrieval results using a sliding window-based bubble sort approach. The window is moved from lower to higher positions, popping the most relevant results to the head, until all the results are ranked. 
% operating on bubble sort principles, to produce global passage rankings. 
This approach, however, cannot be applied to utility-based selection due to the inherent differences. State-of-the-art (SOTA) utility judgment methods are based on pseudo answers that are generated from a group of input documents \cite{zhang2024large,zhang2024iterative}. This requires the student model to also inherit the answer generation capability from the teacher model. Moreover, to ensure decent pseudo-answer quality, results that are more likely to be useful are needed, indicating that the initial passages fed to the model should be of high ranks. 
This argues for a dedicated approach for distilling small utility-based passage selectors and utility judgments of a large number of passages. 
% Crucially, utility-based selection differs from relevance ranking distillation. 
% While relevance distillation relies solely on ranking sequences as training signals, utility-based selection inherently requires pseudo-answer generation to inform passage utility judgments. 
% Distilling this capability therefore necessitates capturing both the teacher LLM's utility judgments and pseudo-answer generation competencies. 
% This dual requirement makes direct application of standard relevance ranking distillation infeasible for utility-based selection. 

% To bridge this gap, we propose a novel utility-based selection distillation approach. 
To this end, we propose a distilling approach that 
% Our approach trains student models to 
jointly learn pseudo-answer generation and utility judgments from teacher LLMs. 
For utility judgments with the student selector on a long initial ranking list, we propose a sliding window method that moves from higher to lower positions. At each step, the selector generates pseudo answers based on the selected useful results, and slides to the next window, which is comprised of the so-far selected useful results and the unseen passages. New selected useful results will be prepended to the selected result pool, and duplicates in the pool will be deleted, maintaining an ordered list of selected useful results. This process is repeated until all the candidate results are judged. 
This process ensures that the final selected useful results are based on the information of the entire candidate results. It also incurs a smaller cost than the above-mentioned ranking distillation due to smaller overlap between windows. 

% After distillation, the student selector employs a front-to-back sliding window strategy to dynamically select optimal passage sets from large-scale candidates, which accounts for the dependency of pseudo-answers on input passages. 

Following the current works for relevance ranking distillation \cite{sun2023chatgpt, pradeep2023rankzephyr,liu2025leveraging}, we also utilize the dataset of 100k queries, sampled from the MS MARCO training set by \cite{sun2023chatgpt} for training. 
We employ Qwen3-32B \cite{qwen3technicalreport} as teacher model to generate both relevance ranking and utility-based selection outputs. 
These outputs are then distilled into Qwen3-1.7B, yielding \rank$_{1.7B}$ (for relevance ranking) and \utility$_{1.7B}$ (for utility-based selection). 
To evaluate RAG performance, we utilize two QA datasets from BEIR \cite{thakur2021beir}: NQ \cite{kwiatkowski2019natural}, and HotpotQA \cite{yang2018hotpotqa}, on two kinds of top-100 initial candidate passages retrieved by two retrievers, i.e., BM25 \cite{robertson2009probabilistic} and BGE-base-en-v1.5 \cite{bge_embedding}.   
Following extensive experimentation, we found that:
(1) For simple questions, such as those in the NQ dataset, relevance ranking (by adjusting various thresholds) demonstrated no statistically significant difference in optimal answer generation performance compared to directly using utility-based selection.
(2) However, for complex questions, exemplified by the HotpotQA dataset, relevance ranking proved insufficient; utility-based selection was more effective in helping large language models (LLMs) identify document sets pertinent to answering the query.
(3) Our utility-based selection method adaptively determines the number of useful passages based on the query and the passages. 
A key consequence is that selecting fewer documents per query enables more unprocessed passages to be handled within each sliding window iteration. 
This results in fewer window iterations for utility-based selection compared to relevance ranking, dramatically reducing the computational cost of LLM inference. 
Using merely 30\% of the computational time, this approach yields higher-quality passages and, consequently, superior answers. 
Additionally, we will release the Qwen3-32B relevance ranking and utility-based selection annotations for the 100k MS MARCO dataset, providing a high-quality dataset for future research in relevance ranking and utility-based selection.

%% file: Sections/Related_work.tex
\section{Related Work} 
In this section, we briefly review existing studies on LLM-based ranking and utility in  Retrieval-Augmented Generation (RAG).
% \subsection{Background} 
% Information retrieval (IR) typically comprises multistage ranking pipelines: retrieval and ranking \cite{guo2022semantic}. 
% Formally, given a query $q$ and a large corpus $\mathcal{C}$ containing numerous documents, the objective of the retrieval stage is to retrieve a list of potentially relevant documents $D = \{d_1, d_2, ..., d_N\}$ from $\mathcal{C}$. 
% Dense retrieval (DR) models are currently the mainstream retrieval method, which encodes the query and documents into dense embeddings, following the dual encoder paradigm. 
% The underpinning backbone of DR has evolved, progressing from encoder-based models \cite{xiao2022retromae, karpukhin2020dense, gao2021condenser, ma2021prop, zhang2025leveraging} and encoder-decoder models \cite{ni2021large} to the contemporary paradigm dominated by decoder-only large language models (LLMs) \cite{zhang2025unleashing, ma2024fine, behnamghader2024llm2vec, li2024llama2vec, springer2024repetition, zeng2025scaling}. 
% Critically, the integration of these large-scale models has further yielded substantial enhancements in retrieval performance. 
% Subsequently, the goal of the ranking stage is to further refine the ordering of the documents within the retrieved list $D$ to optimize ranking performance. 
% Next, we briefly review some existing studies on ranking and Retrieval-Augmented Generation (RAG).  

\subsection{LLM-based Ranking} 
Ranking models frequently utilize a cross-encoder architecture, enabling fine-grained, token-level interactions between the query and the document text \cite{guo2019matchzoo}. 
Ranking approaches are categorized based on how they model the relationship between a query and documents: pointwise \cite{nogueira2019multi, nogueira2020document}, pairwise \cite{pradeep2021expando, burges2005learning}, and listwise \cite{cao2007learning, wang2018lambdaloss, gao2021rethink, zhuang2023rankt5}.  
In the era of pre-trained language models (PLMs), owing to input length constraints, listwise ranking methods are typically implemented using pointwise inputs during training while employing listwise loss functions. 
This approach cannot be considered authentic listwise learning, as it inherently fails to capture query-document interactions during inference. 
The widespread adoption of Large Language Models (LLMs) has spurred significant advancements within the ranking domain, especially zero-shot listwise ranking \cite{liu2025leveraging, reddy2024first, pradeep2023rankzephyr, ren2025self}. 
RankVicuna \cite{pradeep2023rankvicuna} and RankGPT \cite{sun2023chatgpt} pioneered the application of using LLMs for zero-shot listwise ranking. 
RankZephyr \cite{pradeep2023rankzephyr} explored distilling the powerful listwise ranking capabilities of high-performing LLMs into smaller, more efficient models. 
Currently, the primary distillation signal involves relevance ranking sequence generated by LLM. 
Unlike these works, this work explores novel methods to distill the utility-based selection capability of LLMs into smaller models.

\subsection{Utility in RAG}
Retrieval-Augmented Generation (RAG) typically involves two stages: retrieval and generation. While relevance serves as the primary optimization objective for the retriever, the ultimate goal in RAG shifts towards optimizing end-task question answering (QA) performance using outputs from effective retrievers, with less direct emphasis on retrieval metrics themselves \cite{zhang2025leveraging}. Consequently, significant research has focused on the importance of retrieval utility – the measure of a passage's contribution to generating a correct answer – within RAG systems \cite{zhang2024iterative, zhang2025leveraging, zhang2024large, shi2023replug, izacard2023atlas}. \citet{zhang2024large} was the first to propose directly using large language models (LLMs) to judge the utility of retrieved passages, demonstrating that LLM-generated pseudo-answers can effectively assist utility assessment. \citet{zhang2024iterative} further introduced an approach based on Schutz's theory of relevance to iteratively enhance LLM-based utility judgments. However, these methods are constrained by their reliance on utility judgments for only a small number of retrieval candidates per query. 
In contrast, \citet{shi2023replug} proposed quantifying utility score probabilistically, using either the likelihood of the ground-truth answer or performance differences on downstream tasks. This probabilistic approach, however, relies heavily on the availability of ground-truth answers. 
In this work, we extend utility-based selection to a significantly larger scale of retrieval candidates.

%% file: Sections/Method.tex
\section{Utility-based Selection Distillation}  
Unlike relevance emphasizes topical matching between query and passage, utility focuses on the usefulness in generating an answer, which is more important in RAG. 
However, directly using powerful LLMs to judge the utility of large-scale passages is computationally expensive. 
Moreover, ranking approach needs a threshold to determine the number of passages to generate an answer, which is not query-specific. 
Therefore, we propose a novel utility-based selection distillation approach. 
Specifically, utility-based distillation approach trains student selectors to learn pseudo-answer generation and utility judgment from teacher LLMs jointly. 
After distillation, the student model employs a front-to-back sliding window strategy to dynamically select optimal passage sets from large-scale candidates, which accounts for the dependency of pseudo-answers on input passages.

\begin{figure}[h]
  \centering
  \includegraphics[width=\linewidth]{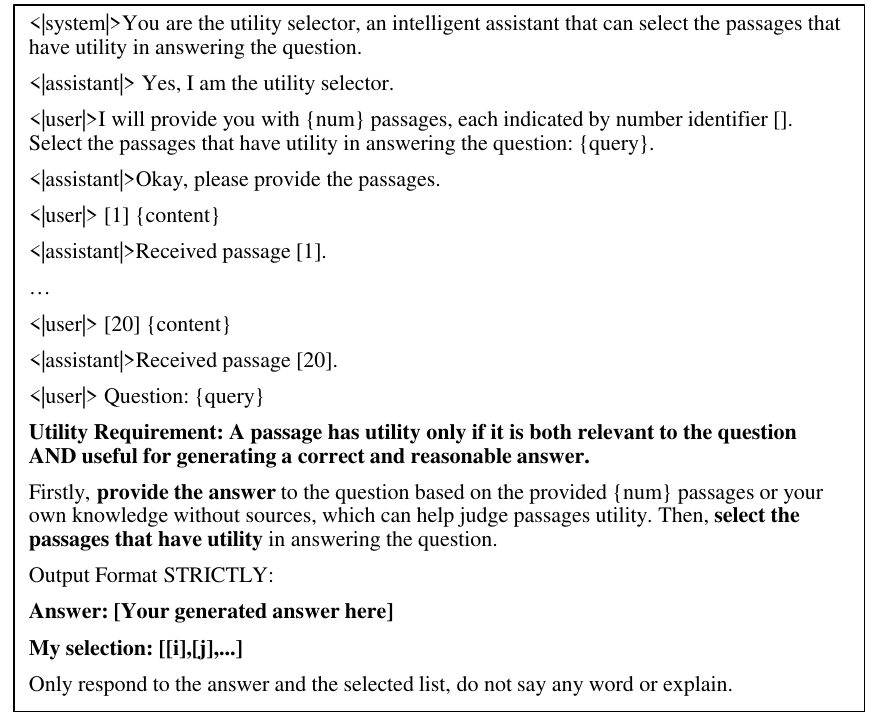}
  \caption{The prompt for utility-based selection. The bold is the special part for utility-based selection compared to relevance ranking.}
  \label{fig:prompt_utility}
\end{figure}

\subsection{Prompt Design} 
Figure \ref{fig:prompt_utility} shows the prompt utilized for utility-based selection. 
The relevance ranking prompt is the same as \citet{sun2023chatgpt}. 

\begin{figure*}[h]
  \centering
  \includegraphics[width=\linewidth]{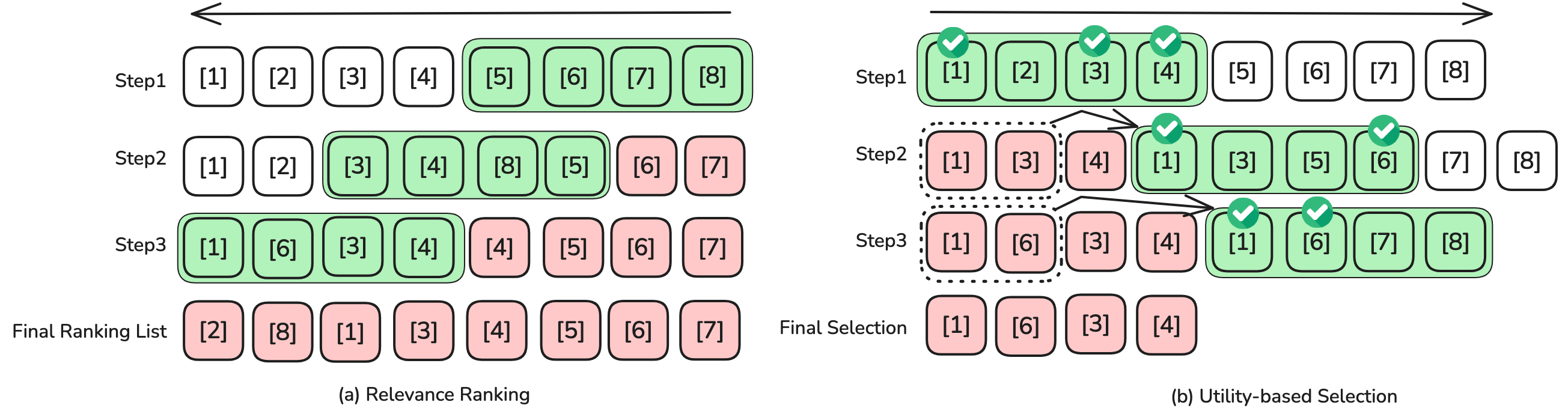}
  \caption{This diagram depicts an 8 passage selection ($M=8$) process using a sliding window $w=4$, stride $s=2$: (a) relevance ranking and (b) utility-based selection. Greed part means the windows, and the arrow indicates the direction of the window sliding. Passages shown in red are those that have undergone re-ranking or preselected passages. White passages denote unprocessed passages. For utility-based selection, $s$ documents (within the dashed-border box) from the preselected queue are placed into the processing window for utility-based selection. Documents selected within the current window are prepended to the head of the preselected queue.}
  \label{fig:sliding}
\end{figure*}
\subsection{Sliding Window for Utility-Based Selection} 
To overcome LLM token limitations during passage re-ranking, prior work often employs a sliding window strategy processed in back-to-front order \cite{sun2023chatgpt}, as shown in Figure \ref{fig:sliding} (a). This approach typically starts re-ranking the last $w$ passages (window size), then slides backwards by stride size $s$, re-ranking passages from position $(M-w-s)$ to $(M-s)$ and repeats until reaching the beginning of the list, where $M$ is the number of candidate passages. This back-to-front progression is effective for relevance permutation generation.

However, utility-based selection presents a distinct challenge: the generation of high-quality pseudo-answers, essential for estimating passage utility, is sensitive to the quality of the input passages provided to the LLM. 
Feeding lower-quality passages in the process can degrade pseudo-answer generation and subsequent utility judgments for all passages. 
To address this dependency and ensure robust pseudo-answer generation, we propose a forward-propagating sliding window strategy, processing passages from front to back.

As depicted in Figure \ref{fig:sliding} (b), our method operates as follows: 
For the first window (positions 1 to $w$), the LLM judges the utility of the $w$ passages within the current window and selects passages that have utility. 
Selected passages are prepended to the preselected queue. 
The original queue content is duplicated and positioned after the new selections.
The window slides forward by a stride size $s$. 
Specifically, the first $s$ passages from the current preselected queue are carried forward as input into the next window, which propagates high-utility context in the sequence. 
The next window now consists of these first $s$ propagated passages from the preselected queue, followed by the next $w-s$ unprocessed passages from the original list (maintaining their sequence). 
The LLM then performs utility-based selection within this new window context. Repeat until all $M$ passages have been processed. The final preselected queue represents the utility-based selection results.

%% file: Sections/Experiments.tex
\section{Experiment Setup}
We distill the relevance ranking and utility-based selection capabilities from larger 32B LLMs into compact 1.7B models, yielding \rank$_{1.7B}$  and \utility$_{1.7B}$. We conduct comparative analyses of both \rank$_{1.7B}$  and \utility$_{1.7B}$  on Retrieval-Augmented Generation (RAG) datasets.  Additionally, we evaluate \rank$_{1.7B}$ in established experiments to compare its ranking performance with prior works \cite{robertson2009probabilistic, bge_embedding, nogueira2019passage, nogueira2020document, sachan2022improving}. 
This section introduces the datasets and implementation details used in this work. 

\subsection{Datasets and Evaluation} 

\heading{RAG}
We use the two subsets of BEIR, i.e., NQ \cite{kwiatkowski2019natural}, which consists of real questions issued to the Google search engine, and HotpotQA \cite{yang2018hotpotqa}, which consists of 7405 QA pairs requiring multi-hop reasoning gathered via Amazon Mechanical Turk. 
We used the queries with ground truth answers from queries on NQ and then filtered 2,255 queries for RAG evaluation.  
We process the top-100 candidate documents retrieved by two retrievers, i.e., BM25 \cite{robertson2009probabilistic} and BGE-base-en-v1.5 \cite{bge_embedding} through distinct pathways:
1) Re-ranking via our \rank$_{1.7B}$ model; 
2) Utility-based selection via our \utility$_{1.7B}$ model. 
Subsequently, the top-$k$ passages from the re-ranked list and all passages selected by the \utility$_{1.7B}$ served as evidence for answer generation using the same underlying generator. 
RAG performance was evaluated on two aspects:
1) \textbf{\textit{Evidence Selection}}: Measured by recall, precision, and micro-F1 score. 
2) \textbf{\textit{Answer Generation}}: Measured by Exact Match (EM) and F1 score.
% RAG evaluation contains selected evidence performance on recall, precision, and F1; answer generation performance on EM and F1. 

\heading{Ranking} We adopt consistent datasets and metrics with \citet{sun2023chatgpt} for ranking performance. 
Specifically, we leveraged test collections from the TREC 2019 and 2020 Deep Learning Tracks \cite{trec_cl19}. 
These tracks employed the MS MARCO v1 passage corpus \cite{bajaj2016ms}, comprising approximately 8.8 million passages. 
To further evaluate \rank's generalization capability beyond the MS MARCO v1 dataset, we assessed it on the BEIR benchmark \cite{thakur2021beir} following \citet{sun2023chatgpt}: TREC-COVID \cite{voorhees2021trec}, NFCorpus \cite{boteva2016full}, Touche \cite{bondarenko2020overview}, DBPedia \cite{hasibi2017dbpedia}, SCIDOCS \cite{cohan2020specter}, SciFact \cite{wadden2020fact}, TREC-News \cite{trec-news}, Roubsust04 \cite{Robust04}. All ranking models re-rank the top-100 passages retrieved by BM25 using pyserini \footnote{\url{https://github.com/castorini/pyserini}} and use nDCG@10  as evaluation metrics. 

\subsection{Implementation Details} 
\heading{LLMs}
We employ the state-of-the-art open-source Qwen3-32B \cite{qwen3technicalreport} model as the teacher for both relevance ranking and utility-based selection functions. 
Through generative distillation, we transfer these capabilities to a compact Qwen3-1.7B student architecture, yielding the specialized models RankQwen$_{1.7B}$ and UtilityQwen$_{1.7B}$. 
In our RAG pipeline, we integrate the distilled models with two distinct generators to examine robustness: Llama-3.1-8B-Instruct \cite{dubey2024llama} and Qwen2.5-7B-Instruct \cite{qwen2.5}. 
To ensure reproducibility, the temperature for all LLMs in this study was set to 0. 

\heading{Teacher Model Annotation}
We employed Qwen2.5-32B as annotators for relevance ranking and utility-based selection. 
The annotated queries were derived from 100K training queries sourced from the MS MARCO v1 passage ranking dataset, originally curated by \citet{sun2023chatgpt}. 
For each query, the top 20 candidate passages were retrieved using BM25 via Pyserini \cite{Lin_etal_SIGIR2021_Pyserini}. 
Qwen3-32B then re-ranked these passages to generate teacher orderings and select passages with utility according to the prompt in Figure \ref{fig:prompt_utility}, which were later distilled into smaller ranker and selector, respectively. 
Following RankZephyr \cite{pradeep2023rankzephyr}, we also enhanced the relevance ranking annotations' quality and robustness by excluding malformed generations from the training data. This exclusion targeted examples exhibiting improper list formatting, missing document identifiers, or repetitive elements. 

\begin{table*}[t]
  \centering
  \caption{Selected evidence performance (\%) and answer generation performance (\%) using relevance ranking (\rank$_{1.7B}$) and utility-based selection (\utility$_{1.7B}$) models with different generators and different Top-100 retrieval results. $^{\dagger}$ means that the performance of answer generation corresponding to the passages selected by Top-$k$ of \rank$_{1.7B}$ is significantly different from that of the passages selected by  \utility$_{1.7B}$ on the paired t-test (p<0.05). ``Llama3.1'' and ``Qwen2.5'' are the ``Llama-3.1-8B-Instruct'' and ``Qwen2.5-7B-Insrtuct'', respectively. \underline{Underline} and \textbf{Bold} indicate the best performance within each group and overall.} 
  \small
   \setlength\tabcolsep{3.5pt}
    \begin{tabular}{llllllllllllll}
    \toprule
    \multicolumn{1}{c}{\multirow{4}[7]{*}{Generator}} & \multicolumn{1}{l}{\multirow{4}[7]{*}{Evidence}} & \multicolumn{6}{c}{Retriever: BGE}            & \multicolumn{6}{c}{Retriever: BM25} \\
\cmidrule(r){3-8}  \cmidrule(r){9-14}  \multicolumn{1}{c}{} & \multicolumn{1}{c}{} & \multicolumn{3}{c}{HotpotQA} & \multicolumn{3}{c}{NQ} & \multicolumn{3}{c}{HotpotQA} & \multicolumn{3}{c}{NQ} \\
\cmidrule(r){3-5}  \cmidrule(r){6-8}  \cmidrule(r){9-11} \cmidrule(r){12-14}     \multicolumn{1}{c}{} &       & Evidence & \multicolumn{2}{c}{Answer} & Evidence & \multicolumn{2}{c}{Answer} & Evidence & \multicolumn{2}{c}{Answer} & Evidence & \multicolumn{2}{c}{Answer} \\
\cmidrule(r){3-3} \cmidrule(r){4-5}   \cmidrule(r){6-6} \cmidrule(r){7-8}  \cmidrule(r){9-9} \cmidrule(r){10-11} \cmidrule(r){12-12} \cmidrule(r){13-14}     \multicolumn{1}{c}{} &       & Micro-F1 & EM    & F1    & Micro-F1 & EM    & F1    & Micro-F1 & EM    & F1    & Micro-F1 & EM    & F1 \\
    \midrule
    % \multicolumn{1}{r}{\multirow{9}[17]{*}{llama3.1}} 
    \multicolumn{1}{c}{\multirow{9}{*}{Llama3.1}}
& \rank$_{1.7B}$ (Top-5) & \underline{41.70}  & \underline{38.39}$^\dagger$  &\underline{ 49.68}$^\dagger$   & \underline{28.44 } & \underline{48.03}  & \underline{59.93}  & \underline{38.48}  & \underline{37.81}$^\dagger$  & \underline{48.97}$^\dagger$  & \underline{24.14}  & \underline{\textbf{45.32}}  & \underline{56.61}  \\
& \rank$_{1.7B}$ (Top-10) & 34.55  & 36.92$^\dagger$   & 48.21$^\dagger$   & 23.52  & 47.72$^\dagger$   & 59.34$^\dagger$   & 31.84  & 36.83$^\dagger$  & 47.91$^\dagger$  & 19.71  & 45.01  & 56.27  \\
& \rank$_{1.7B}$ (Top-15) & 29.93  & 36.81$^\dagger$   & 48.08$^\dagger$   & 20.31  & 47.72$^\dagger$   & 59.23$^\dagger$   & 29.29  & 36.45$^\dagger$  & 47.65$^\dagger$  & 16.87  & 45.14  & 56.22  \\
& \rank$_{1.7B}$ (Top-20) & 26.58  & 36.43$^\dagger$   & 47.49$^\dagger$   & 17.97  & 46.92$^\dagger$   & 58.30$^\dagger$   & 24.64  & 35.84$^\dagger$  & 46.84$^\dagger$  & 14.84  & 44.61  & 55.35$^\dagger$  \\
& \rank$_{1.7B}$ (Top-40) & 23.31  & 35.33$^\dagger$   & 46.02$^\dagger$   & 15.67  & 46.47$^\dagger$   & 58.16$^\dagger$   & 20.80  & 35.69$^\dagger$  & 46.36$^\dagger$  & 12.89  & 43.50  & 54.27$^\dagger$  \\
& \rank$_{1.7B}$ (Top-60) & 20.67  & 34.63$^\dagger$   & 45.29$^\dagger$   & 13.83  & 46.12$^\dagger$   & 57.43$^\dagger$   & 17.97  & 34.31$^\dagger$  & 44.90$^\dagger$  & 11.35  & 41.55$^\dagger$  & 52.39$^\dagger$  \\
& \rank$_{1.7B}$ (Top-80) & 18.55  & 33.71$^\dagger$   & 44.29$^\dagger$   & 12.36  & 45.35$^\dagger$   & 56.01$^\dagger$   & 15.83  & 33.54$^\dagger$  & 43.99$^\dagger$  & 10.13  & 41.02$^\dagger$  & 51.90$^\dagger$  \\
& \rank$_{1.7B}$ (Top-100) & 16.83  & 32.92$^\dagger$   & 43.62$^\dagger$   & 11.18  & 45.59$^\dagger$   & 56.53$^\dagger$  & 14.16  & 32.75$^\dagger$  & 43.09$^\dagger$  & \phantom{1}9.15  & 38.36$^\dagger$  & 49.57$^\dagger$  \\
\cmidrule{2-14}          & \utility$_{1.7B}$ & \textbf{60.58}  & \textbf{41.59}  & \textbf{53.56} & \textbf{30.54}  & \textbf{49.31}  & \textbf{60.86}   & \textbf{57.03}  & \textbf{39.00}  & \textbf{50.57}   & \textbf{28.80}  & 45.06  & \textbf{56.83}  \\
\midrule
% \multicolumn{1}{r}{\multirow{9}[16]{*}{Qwen2.5}} 
\multicolumn{1}{c}{\multirow{9}{*}{Qwen2.5}} 
& \rank$_{1.7B}$ (Top-5) & \underline{41.70}  & 42.13$^\dagger$  & 53.98$^\dagger$  & \underline{28.44}  & 48.03  & 60.17  & \underline{38.48}  & \underline{39.97}$^\dagger$  & \underline{51.53}$^\dagger$  & \underline{24.14}  & 44.88  & 56.61  \\
& \rank$_{1.7B}$ (Top-10) & 34.55  & \underline{42.36}$^\dagger$  & \underline{54.15}$^\dagger$  & 23.52  & \underline{48.12}  & 60.54  & 31.84  & 39.69$^\dagger$  & 50.98$^\dagger$  & 19.71  & \underline{\textbf{44.97}}  & \underline{\textbf{57.07}}  \\
& \rank$_{1.7B}$ (Top-15) & 29.93  & 42.27$^\dagger$  & 54.00$^\dagger$  & 20.31  & 47.85  & \underline{60.56}  & 29.29  & 39.32$^\dagger$  & 50.70$^\dagger$  & 16.87  & 44.70  & 56.85  \\
& \rank$_{1.7B}$ (Top-20) & 26.58  & 42.21$^\dagger$  & 53.86$^\dagger$  & 17.97  & 47.89  & 60.55  & 24.64  & 39.58$^\dagger$  & 50.87$^\dagger$  & 14.84  & 44.52  & 56.77  \\
& \rank$_{1.7B}$ (Top-40) & 23.31  & 42.47$^\dagger$  & 54.25$^\dagger$  & 15.67  & 47.01  & 59.80  & 20.80  & 39.43$^\dagger$  & 50.80$^\dagger$  & 12.89  & 44.21  & 56.46  \\
& \rank$_{1.7B}$ (Top-60) & 20.67  & 41.61$^\dagger$  & 53.48$^\dagger$  & 13.83  & 47.23  & 59.55$^\dagger$  & 17.97  & 38.76$^\dagger$  & 49.87$^\dagger$  & 11.35  & 43.55  & 55.69  \\
& \rank$_{1.7B}$ (Top-80) & 18.55  & 41.22$^\dagger$  & 52.98$^\dagger$  & 12.36  & 46.25$^\dagger$  & 58.72$^\dagger$  & 15.83  & 38.12$^\dagger$  & 49.14$^\dagger$  & 10.13  & 43.10  & 55.17$^\dagger$  \\
& \rank$_{1.7B}$ (Top-100) & 16.83  & 40.53$^\dagger$  & 52.21$^\dagger$  & 11.18  & 46.74$^\dagger$  & 58.95$^\dagger$  & 14.16  & 37.42$^\dagger$  & 48.24$^\dagger$  & \phantom{1}9.15  & 42.26$^\dagger$  & 53.79$^\dagger$  \\
\cmidrule{2-14} & \utility$_{1.7B}$ & \textbf{60.58}  & \textbf{43.59}  & \textbf{55.73} & \textbf{30.54}  & \textbf{48.07}   & \textbf{60.95}  & \textbf{57.03}  & \textbf{41.00}  & \textbf{52.57}  & \textbf{28.80}  & 44.61 & 57.00 \\
\bottomrule   
\end{tabular}%
  \label{tab:ranking_vs_selection}%
\end{table*}%

\heading{Distillation Training Details}
Using the axolotl library\footnote{\url{https://github.com/OpenAccess-AI-Collective/axolotl}} on eight NVIDIA A800 80GB GPUs, we trained the 1.7B parameter Qwen 3 model \cite{qwen3technicalreport} for three epochs on both ranking and selection data generated by Qwen3-32B. 
Training employed bfloat16 precision, an effective batch size of 64, and a learning rate of \(5 \times 10^{-6}\).  
Mirroring RankZephyr \cite{pradeep2023rankzephyr}, we also incorporated noisy embeddings, a method demonstrated to enhance instruction fine-tuning \cite{jain2023neftune}. 

\heading{Inference Details}
For both relevance ranking and utility-based selection tasks, we employed two retrievers, BM25 \cite{robertson2009probabilistic} and BGE-base-en-v1.5 \cite{bge_embedding}, to retrieve the initial top-100 ($M=100$) candidate lists. 
During inference, two sliding window strategies were applied to both tasks, respectively. 
The window size was set to 20 ($w=20$), with a stride size of 10 ($s=10$).

\section{Experimental Results}
We propose a novel utility-based selection distillation approach to transfer the capability for utility judgment into smaller, more efficient models,  and conducted a comparative analysis between relevance ranking (top-$k$ passages are used to answer generation) and utility-based selection for answer generation on RAG. 
The RAG performance, presented comprehensively in Table \ref{tab:ranking_vs_selection}, reveal four key insights:
\begin{enumerate*}[label=(\roman*)]
\item Evidence performance:  Our utility-based selection achieves best performance among different generators and different initial retrieval candidate lists. 
\item Answer generation performance:  For simple questions, such as those in the NQ dataset, relevance ranking (by adjusting various thresholds) demonstrated no statistically significant difference in optimal answer generation performance compared to directly using utility-based selection. However, for complex questions, exemplified by the HotpotQA dataset, relevance ranking proved insufficient; utility-based selection was more effective in helping large language models (LLMs) identify document sets pertinent to answering the query. This substantial improvement stems from the inherent requirement of multi-hop reasoning: generating an answer often necessitates identifying a set of complementary passages, i.e., set-based utility-based selection, that collectively provide the necessary evidence, which is beyond relevance. 
\item With a better initial retrieval candidate, such as BGE, the \utility$_{1.7B}$ model shows a larger performance margin compared to \rank$_{1.7B}$. This improvement may be because better initial retrieval candidates enable \utility to generate higher-quality pseudo-answers, which further enhances its utility judgment performance. 
\item Given the same re-ranked results, Llama-3.1-8B-Instruct achieves peak performance using only the top 5 passages. Conversely, Qwen2.5-7B-Instruct performs optimally across various top-$k$ values, indicating potentially distinct passage digestion capabilities. 
Crucially, utility-based selection demonstrates superior robustness. 
It consistently outperforms fixed top-$k$ approaches across diverse generators, datasets, and initial retrieval lists. 
Therefore, our approach provides a robust alternative by dynamically selecting the most useful passages. It achieves answer quality comparable to or exceeding the best-performing fixed top-$k$ cutoff for any given configuration, while entirely eliminating the need for manual $k$-selection. 
\end{enumerate*}
These findings demonstrate that \utility$_{1.7B}$ offers a superior and more adaptable paradigm for passage selection in RAG, particularly for complex information needs, by explicitly optimizing for the collective utility of passages in supporting accurate answer generation.

\begin{table*}[t]
  \centering
  \caption{Results (nDCG@10) (\%) on TREC and BEIR. All ranking models re-rank the top-100 BM25 passages. \underline{Underline} and \textbf{Bold} indicate the best performance within each group and overall. $^*$ means the results are copied from the original paper.}
  \small
   \setlength\tabcolsep{3.5pt}
    \begin{tabular}{ll|l|rrrrrrrrr}
    \toprule
    Method & \multicolumn{1}{l|}{DL19} & \multicolumn{1}{l}{DL20} & \multicolumn{1}{|l}{Covid} & \multicolumn{1}{l}{NFCorpus} & \multicolumn{1}{l}{Touche} & \multicolumn{1}{l}{DBPedia} & \multicolumn{1}{l}{SciFact} & \multicolumn{1}{l}{Signal} & \multicolumn{1}{l}{News} & \multicolumn{1}{l}{Robust04} & \multicolumn{1}{l}{Avg} \\
    \midrule
    BM25$^*$ \cite{robertson2009probabilistic}  & 50.58 & 47.96 & 59.47 & 30.75 & \underline{44.22} & 31.80  & 67.89 & \underline{33.05} & 39.52 & 40.70  & 43.42 \\
    BGE-base-en-v1.5 \cite{bge_embedding} &  \underline{70.60} &  \underline{\textbf{71.70}} & \underline{78.10} & \underline{37.30}  & 25.70  &  \underline{40.70}  &  \underline{74.10} &  28.90 & \underline{44.20}  &  \underline{44.70} & \underline{51.60}  \\
    \midrule
    monoBERT(340M)$^*$  \cite{nogueira2019passage} & 70.50  & 67.28 & 70.01 & 36.88 & 31.75 & 41.87 & 71.36 & 31.44 & 44.62 & 49.35 & 47.16 \\
    monoT5(220M)$^*$  \cite{nogueira2020document} & 71.48 & 66.99 & 78.34 & 37.38 & 30.82 & 42.42 & 73.40  & 31.67 & 46.83 & 51.72 & 49.07 \\
    monoT5(3B)$^*$  \cite{nogueira2020document} & 71.83 & \underline{68.89} & 80.71 & \underline{\textbf{38.97}} & 32.41 & \underline{44.45} & \underline{76.57} & \underline{32.55} & \underline{48.49} & \underline{56.71} & \underline{51.36} \\
    Cohere Rerank-v2$^*$    & \underline{73.22} & 67.08 & \underline{81.81} & 36.36 & \underline{32.51} & 42.51 & 74.44 & 29.60  & 47.59 & 50.78 & 49.45 \\
    UPR(FLAN-T5-XL)$^*$  \cite{sachan2022improving} & 53.85 & 56.02 & 68.11 & 35.04 & 19.69 & 30.91 & 72.69 & 31.91 & 43.11 & 42.43 & 42.99 \\
    \midrule
    LLM Zero-shot Inference (Permutation generation)   \\
    \midrule
    ChatGPT$^*$  \cite{sun2023chatgpt} & 65.80  & 62.91  & 76.67  & 35.62  & 36.18  & 44.47  & 70.43  & 32.12  & 48.85  & 50.62  & 49.37  \\
    % \midrule
    GPT-4$^*$  \cite{sun2023chatgpt}  & \underline{\textbf{75.59}}  & \underline{70.56}  & \underline{\textbf{85.51}}  & 38.47  & 38.57  & \underline{\textbf{47.12}}  & 74.95  & \underline{\textbf{34.40}}  & \underline{\textbf{52.89}}  & 57.55  & \underline{\textbf{53.68}}  \\
    Qwen3$_{1.7B}$ & 60.32 & 55.83&	65.65	& 33.61 & \underline{\textbf{45.47}} &	39.05	& 64.36	& 32.53 &	43.29 &	40.44	& 45.55  \\
    Qwen3$_{32B}$ &73.33 &	70.35&	85.48	& \underline{38.77}	&32.27&	44.97&	 \underline{\textbf{77.71}} &	32.08 &	50.67 &	\underline{\textbf{60.65}} & 52.82 \\
    \midrule
    LLM Distillation Training  \\
    
    % \midrule
    % LLM Inference(Permutation Distillation) &       &       &       &       &       &       &       &       &       &       &  \\
    % RankVicuna-7b & 71.80 & 66.89 &78.32 &36.87& 31.81 &45.40 &74.23 &34.28&51.13 & 52.91 &50.62 \\
    % \midrule
    % RankQwen3 1.7B(Qwen3 32B w/ think) & 70.89  & 68.83  & 85.49  & 38.82  & 37.63  & 41.76  & 74.31  & 32.98  & 48.04  & 56.93  & 55.57  \\
    \midrule
    % RankVicuna &    66.82   &    65.49   &  -     &       &       &       &       &       &       &       &  \\
    RankMistral$_{7B}$$^*$   \cite{liu2025leveraging} &  \underline{71.73} &  68.07 & 78.00	& 33.10 &	27.46 &	37.71 &	66.22 &	30.04	& 37.10	&39.54	& 43.65\\
    PE-Rank$_{7B}$$^*$  \cite{liu2025leveraging} &70.48 &63.54& 77.72 &  36.39&  33.06&  40.05&  69.38 & 33.74 & \underline{49.70} & 47.40 & 48.43 \\
    \rank$_{1.7B}$ (Teacher: Qwen3-32B $w/$ think) & 70.89  & 68.83  & \underline{85.49}  & \underline{38.82}  & 37.63  & 41.76  & 74.31  & 32.98  & 48.04  & 56.93  & 52.00 \\
    \rank$_{1.7B}$ (Teacher: Qwen3-32B $w/o$ think) & 71.36  & \underline{69.13}  & 83.61  & 38.19  & \underline{38.11}  & \underline{42.27}  & \underline{74.64}  & \underline{33.87}  & 48.56  & \underline{57.64}  & \underline{52.11}  \\
    \bottomrule
    \end{tabular}%
  \label{tab:ranking}%
\end{table*}%

%% file: Sections/Futher_analyse.tex
\section{Further Analysis}

We further analyze the (1) ranking performance of \rank$_{1.7B}$ used in our experiments, (2) the impact of distillation training and the LLM backbone, (3) a statistics study on utility-based selection, and (4) the inference efficiency of both relevance ranking and utility-based selection. 
\begin{table*}[t]
  \centering
  % \small
  \setlength\tabcolsep{3.0pt}
  \caption{Different evidence selection and answer generation performance (\%) on the top-100 BM25 retrieval results using utility-based selection approach. The generator is the Llama-3.1-8B-Instruct. \textbf{Bold} represents the best performance.}
    \begin{tabular}{lrrrrrrrrrr}
    \toprule
    \multirow{3}[6]{*}{Model} & \multicolumn{5}{c}{HotpotQA}         & \multicolumn{5}{c}{NQ} \\
\cmidrule(r){2-6}    \cmidrule{7-11}       & \multicolumn{3}{c}{Evidence} & \multicolumn{2}{c}{Answer} & \multicolumn{3}{c}{Evidence} & \multicolumn{2}{c}{Answer} \\
\cmidrule(r){2-4}    \cmidrule(r){5-6}     \cmidrule(r){7-9}    \cmidrule(r){10-11}              & Recall & Precision & Micro-F1    & EM    & F1    & Recall & Precision & Micro-F1     & EM    & F1 \\
    \midrule
    Qwen3-32B (Teacher Model) & \textbf{71.37}  & \textbf{69.29}  & \textbf{70.32}  & \textbf{42.55}  & \textbf{54.63}  & \textbf{68.74}  & \textbf{21.50}  & \textbf{32.76}  & \textbf{46.25}  & \textbf{58.74}  \\
    Qwen3-1.7B (Student Model $w/o$ Distillation)  & 51.68  & 39.61  & 44.85  & 33.09  & 43.88  & 61.42  & \phantom{1}9.47  & 16.41  & 41.91  & 52.94  \\
    \midrule
    \utility$_{1.7B}$ (Student Model $w/$ Distillation) & 58.68  & 55.47  & 57.03  & 39.00  & 50.57  & 67.20  & 18.32  & 28.80  & 45.06  & 56.83  \\
    \bottomrule
    \end{tabular}%
  \label{tab:distillation}%
\end{table*}%

\subsection{Ranking Performance}
To validate the ranking efficacy of our \rank$_{1.7B}$, we comprehensively compare different ranking model performance across the TREC and BEIR datasets, comprising:
\begin{enumerate*}
\item Retrievers: BM25  \cite{robertson2009probabilistic}  and BGE-base-en-v1.5 \cite{bge_embedding}; 
\item Supervised cross-encoders: monoBERT \cite{nogueira2019passage}, monoT5 \cite{nogueira2020document}, and Cohere Rerank-v2\footnote{\url{https://cohere.com/rerank}}; 
\item Unsupervised cross-encoder: UPR \cite{sachan2022improving}; 
\item Zero-shot LLM listwise re-ranking;  
\item Relevance ranking distillation, 
\end{enumerate*}
as shown in Table \ref{tab:ranking}.   
We can observe that:
\begin{enumerate*}[label=(\roman*)]
\item Zero-Shot LLM Superiority: Large language models (LLMs) operating in a zero-shot listwise ranking paradigm (e.g., GPT-4) demonstrably outperform supervised cross-encoder baselines (e.g., monoT5-3B) on both the TREC and BEIR benchmark, indicating the profound semantic understanding capabilities inherent in LLMs, enabling highly effective relevance ranking without task-specific training. 
\item Our distilled model, \rank$_{1.7B}$, notably surpasses the  ChatGPT by 5.5\% in terms of nDCG@10 on BEIR average. 
\rank$_{1.7B}$ establishes a new state-of-the-art ranker among open source distillation efficient listwise rankers and highlights the exceptional ranking aptitude of the Qwen3 backbone upon which \rank$_{1.7B}$ is built. 
\item \textbf{\texttt{Discussion on Reasoning LLMs for Ranking}}: 
Qwen3-series models support an optional reasoning process (thinking) during generation. We conducted relevance ranking annotations under two configurations: 1) Qwen3-32B with reasoning process and 2) Qwen3-32B without reasoning process. For efficiency, we distilled only the final ranking sequences from both configurations into the student ranker, rather than simulating intermediate reasoning steps. Both distilled rankers achieved better ranking performance compared to other rankers. The non-reasoning ranking sequences distillation slightly outperformed the reasoning variant. Based on this efficiency-performance tradeoff, all LLMs in this study operate without reasoning processes unless explicitly noted. 
\end{enumerate*} 
In summary, these results provide a robust, high-performance ranking foundation essential for the subsequent in-depth analysis in this work.

\subsection{Impact of Distillation Training}
To validate the effectiveness of distillation training, we performed utility-based selection on the top-100 BM25 retrieval results for NQ and HotpotQA directly using both the teacher model (Qwen3-32B) and the student model (Qwen3-1.7B) without distillation training. 
The experimental results are summarized in Table \ref{tab:distillation}. We observe that the student model fine-tuned via generation distillation achieves a substantial improvement in both the quality of the selected passages list and the answer generation performance. 

\begin{table}[htbp]
  % \centering
  \small
  \setlength\tabcolsep{2.5pt}
   \caption{Different evidence selection and answer generation performance (\%) on the top-100 BGE retrieval results of the HotpotQA dataset. The generator is the Llama-3.1-8B-Instruct. ``Ranking ($k$)" means the top-$k$ re-ranked (from re-ranked top-100 first retrieval results) documents are used to answer questions. \underline{Underline} and \textbf{Bold} indicate the best performance within each group and overall. ``F1'' in evidence evaluation is ``Micro-F1''.}
    \begin{tabular}{lllllll}
    \toprule
     \multirow{2}[4]{*}{LLM} & \multirow{2}[4]{*}{Method} & \multicolumn{3}{c}{Evidence} & \multicolumn{2}{c}{Answer} \\
\cmidrule(r){3-5}   \cmidrule(r){6-7}         &       & Recall & Precision & F1    & EM    & F1 \\
    \midrule
    \multirow{8}[2]{*}{\rank$_{1.7B}$} &  Ranking (5) & 72.97  & \underline{29.19}  & \underline{41.70}  & \underline{38.39}  & \underline{49.68}  \\
          & Ranking (10) & 75.52  & 22.40  & 34.55  & 36.92  & 48.21  \\
          & Ranking (15) & 77.46  & 18.55  & 29.93  & 36.81  & 48.08  \\
          & Ranking (20) & 78.83  & 15.98  & 26.58  & 36.43  & 47.49  \\
          & Ranking (40) & 80.09  & 13.64  & 23.31  & 35.33  & 46.02  \\
          & Ranking (60) & 81.10  & 11.84  & 20.67  & 34.63  & 45.29  \\
          & Ranking (80) & 81.91  & 10.46  & 18.55  & 33.71  & 44.29  \\
          & Ranking (100) & \underline{82.55}  &  \phantom{1}9.37  & 16.83  & 32.92  & 43.62  \\
    \midrule
    \multirow{8}[2]{*}{Qwen3-32B} &  Ranking (5) & 80.91  & \underline{32.36}  & \underline{46.24}  & \underline{40.12}  & \underline{52.03}  \\
          & Ranking (10) & 82.10  & 24.51  & 37.75  & 39.07  & 50.70  \\
          & Ranking (15) & 82.83  & 20.09  & 32.33  & 38.26  & 49.61  \\
          & Ranking (20) & 83.31  & 17.18  & 28.49  & 38.07  & 49.39  \\
          & Ranking (40) & 83.85  & 14.61  & 24.88  & 37.16  & 48.44  \\
          &  Ranking (60) & 84.30  & 12.65  & 22.00  & 36.15  & 47.12  \\
          &  Ranking (80) & 84.67  & 11.16  & 19.71  & 34.83  & 45.86  \\
          &  Ranking (100) & \underline{\textbf{84.96}}  & \phantom{1}9.98  & 17.86  & 34.17  & 45.24  \\
    \midrule
    \utility$_{1.7B}$ & Selection & 63.73  & 57.73  & 60.58  & 41.59  & 53.56  \\
    Qwen3-32B & Selection & 78.56  & \textbf{72.18}  & \textbf{75.24}  & \textbf{46.39}  & \textbf{59.20}  \\
    \bottomrule
    \end{tabular}%
  \label{tab:llm_backbone}%
\end{table}%

% \begin{table}[t]
%   \centering
%   \small
%    % \renewcommand{\arraystretch}{0.95}
%    \setlength\tabcolsep{3.5pt}
%   \caption{Different evidence selection and answer generation performance (\%) on the top-100 BGE retrieval results. The generator is Llama-3.1-8B-Instruct. \textbf{Bold} represents the best performance.}
%     \begin{tabular}{llllll}
%     \toprule
%     \multicolumn{1}{c}{\multirow{2}[4]{*}{Method}} & \multicolumn{3}{c}{Evidence} & \multicolumn{2}{c}{Answer} \\
% \cmidrule{2-6}          & \multicolumn{1}{l}{Recall} & \multicolumn{1}{l}{Precision} & \multicolumn{1}{l}{F1} & \multicolumn{1}{l}{EM} & \multicolumn{1}{l}{F1} \\
%     \midrule
%     Relevance Ranking(top-5) & 80.91  & 32.36  & 46.24  & 40.12  & 52.03  \\
%     Relevance Ranking(top-10) & 82.10  & 24.51  & 37.75  & 39.07  & 50.70  \\
%     Relevance Ranking(top-15) & 82.83  & 20.09  & 32.33  & 38.26  & 49.61  \\
%     Relevance Ranking(top-20) & 83.31  & 17.18  & 28.49  & 38.07  & 49.39  \\
%     Relevance Ranking(top-40) & 83.85  & 14.61  & 24.88  & 37.16  & 48.44  \\
%     Relevance Ranking(top-60) & 84.30  & 12.65  & 22.00  & 36.15  & 47.12  \\
%     Relevance Ranking(top-80) & 84.67  & 11.16  & 19.71  & 34.83  & 45.86  \\
%     Relevance Ranking(top-100) & 84.96  & 9.98  & 17.86  & 34.17  & 45.24  \\
%     \midrule
%     utility-based selection & 78.56  & \textbf{72.18}  & \textbf{75.24}  & \textbf{46.39}  & \textbf{59.20}  \\
%     \bottomrule
%     \end{tabular}%
%   \label{tab:llm_backbone}%
% \end{table}%
\begin{figure}[h]
  \centering
  \includegraphics[width=\linewidth]{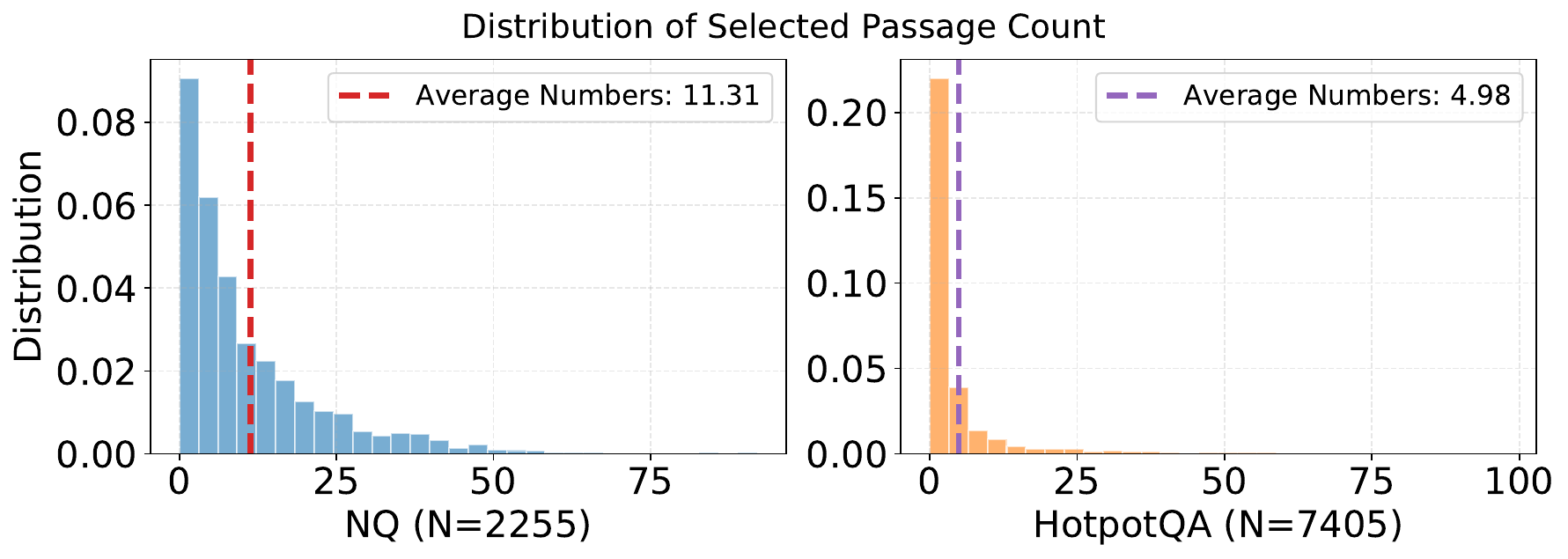}
  \caption{Distribution of selected passage count of \utility$_{1.7B}$ on the NQ and HotpotQA datasets with BM25 retrieval results. ``N'' means the query numbers of the datasets.}
  % \Description{A woman and a girl in white dresses sit in an open car.}
  \label{fig:distribution}
\end{figure}

\begin{figure*}[h]
  \centering
  \includegraphics[width=\linewidth]{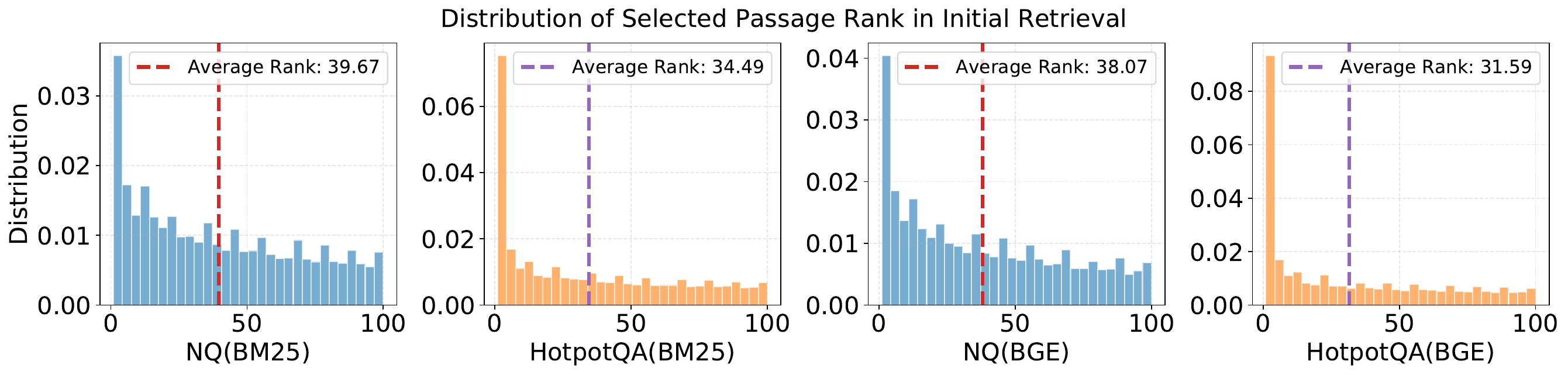}
  \caption{Distribution of selected passage rank in initial retrieval of \utility$_{1.7B}$ on the NQ and HotpotQA datasets with initial BM25/BGE retrieval results.}
  % \Description{A woman and a girl in white dresses sit in an open car.}
  \label{fig:rank_distribution}
\end{figure*}

\subsection{Impact of LLM Backbone}
To evaluate the performance ceiling of relevance ranking versus utility-based selection in our experimental setup, we directly employed the teacher model (Qwen3-32B) to perform both relevance ranking and utility-based selection on the top-100 BGE-base-en-v1.5 retrieval results for HotpotQA. The selected document sets were then used for downstream answer generation, with results presented in Table \ref{tab:llm_backbone}. Our key observations are: 
1) utility-based selection using the teacher model significantly outperforms threshold-based document selection via relevance ranking in terms of answer generation quality. This further underscores the efficacy of the utility-based approach for evidence selection. 
2) The performance gap between utility-based selection and relevance ranking observed with Qwen3-32B is more pronounced than the corresponding gap between \utility$_{1.7B}$ and \rank$_{1.7B}$. 
This suggests that utility-based selection benefits more substantially from stronger LLM backbones, indicating its greater potential for improvement in RAG applications compared to traditional ranking methods.

\subsection{Statistics of Utility-Based Selection}
Figure \ref{fig:distribution} shows the distribution of passage counts selected as useful by \utility$_{1.7B}$ from the top-100 BM25 results. 
\utility$_{1.7B}$ selects a varying number of useful passages across different queries, with significantly more passages chosen for NQ queries.

We further analyze the distribution of selected passage ranks from the initial retrieval, as shown in Figure \ref{fig:rank_distribution}. Key observations include:
(1) Our utility-based selection approach retrieves passages across diverse initial ranks, indicating that useful passages may appear at varying positions. This suggests that a fixed threshold may be suboptimal for handling different queries.
(2) The average rank of selected passages is smaller with BGE than with BM25, implying that improved initial ranking (e.g., via BGE) positions utility passages closer to the top.

% \subsection{Sliding Window for utility-based selection}
\subsection{Inference Efficiency}
% \begin{table}[htbp]
%   \centering
%   \small
%   \setlength\tabcolsep{3pt}
%   \caption{Analysis of inference efficiency on NQ and HotpotQA datasets with Top-100 BGE retrieval results. ``Avg.win\_num'' represents the average number of windows for each query during the sliding window process. For fair comparison, all the experiments are conducted on eight NVIDIA A800 80GB GPUs.}
%     \begin{tabular}{lllllll}
%     \toprule
%     \multicolumn{1}{c}{\multirow{2}[4]{*}{Method}} & \multicolumn{2}{l}{Avg.win\_num} & \multicolumn{2}{c}{Time} & \multicolumn{2}{c}{RAG(F1)}\\
% \cmidrule(r){2-3} \cmidrule(r){4-5} \cmidrule(r){6-7}    \multicolumn{1}{c}{} & \multicolumn{1}{l}{HotpotQA} & \multicolumn{1}{l}{NQ} & \multicolumn{1}{l}{HotpotQA} & NQ    & \multicolumn{1}{l}{HotpotQA} & \multicolumn{1}{l}{NQ} \\
%     \midrule
%     RankQwen$_{1.7B}$ & 9     & 9    & 11.2h & 3.4h    & 54.25 & 60.17 \\
%     UtilityQwen$_{1.7B}$ & 6     & 7    &   \phantom{1}3.4h    &  1.2h     & 55.73 & 60.95 \\
%     Qwen3-32B (Ranking) &      &     &   &   &  &  \\
%     Qwen3-32B (Selection) &      &     &   &   &  &  \\
%     \bottomrule
%     \end{tabular}%
%   \label{tab:efficiency}%
% \end{table}%

\begin{table}[t]
  \centering
  \caption{Analysis of inference efficiency on the HotpotQA datasets with Top-100 BGE retrieval results. ``Avg.win\_num'' represents the average number of windows for each query during the sliding window process. For fair comparison, all the experiments are conducted on eight NVIDIA A800 80GB GPUs. The generator is the Llama-3.1-8B-Instruct.}
    \begin{tabular}{lrrr}
    \toprule
    \multicolumn{1}{l}{Method} & Avg.win\_num &  Time & Answer-F1 \\
    \midrule
    Qwen3-32B (Ranking) & 9.0     & 23.3h & 52.03 \\
    Qwen3-32B (Selection) & 6.1   & \phantom{1}6.9h & 59.20 \\
    RankQwen$_{1.7B}$ & 9.0     & 11.2h & 49.68 \\
    UtilityQwen$_{1.7B}$ & 6.4 & \phantom{1}3.4h  & 53.36 \\
    \bottomrule
    \end{tabular}%
 \label{tab:efficiency}%
\end{table}%

Inference efficiency of large language models (LLMs) is a critical practical consideration in RAG deployments. 
Even after distilling the ranking or selection capabilities from a 32B model to a 1.7B model, the inference latency remains significantly higher than that of a 110M cross-encoder model. 
Therefore, we conducted a comparative analysis of the inference efficiency between \rank$_{1.7B}$ and \utility$_{1.7B}$. 
As shown in the Table \ref{tab:efficiency}, we can observe that: 
\begin{enumerate*}
    \item \textbf{Distillation necessity}: Qwen3-32B incurs significantly higher costs than Qwen3-1.7B, underscoring the importance of distillation for practical deployment.  
    \item  \textbf{Efficiency of utility-based selection}: 
    This approach significantly reduces context window consumption and achieves approximately 30\% lower inference latency than relevance ranking.
    Notably, even Qwen3-32B for utility-based selection requires less computational cost than \rank$_{1.7B}$. 
    Furthermore, utility-based selection demonstrates more stable performance compared to the relevance ranking approach. 
\end{enumerate*}

%% file: Sections/Conclusion.tex
\section{Conclusion} 
Retrieval-augmented generation (RAG) demands a paradigm shift from relevance to utility—where passages are valued not merely for topical alignment but for their usefulness to enable complete, accurate answers. 
Utility-based RAG faces a critical scalability bottleneck: the high cost of using LLMs for utility judgments limits practical deployment to $\sim$20 passages per query, significantly hindering performance over large-scale passages.   
Since ranking requires a fixed strategy to determine passage count, we propose distilling the utility-based selection capability from larger models into smaller, efficient models. 
Key contributions and findings:
(1) We propose a novel utility-based selection distillation method that trains efficient student models (e.g., Qwen3-1.7B) to jointly learn pseudo-answer generation and utility assessment from teacher LLMs (Qwen3-32B). This bypasses the need for direct LLM inference during deployment. 
(2) We empirically validate that utility-driven passage selection (unlike traditional relevance ranking) is essential for complex QA tasks (e.g., HotpotQA), where synthesizing information across multiple documents is paramount. 
(3) Our distilled model employs a front-to-back sliding window strategy to adaptively select minimal high-utility passage sets from large-scale candidates (top-100), dynamically skipping low-utility regions. 
Adaptive selection yields higher-quality answers with fewer passages, accelerating inference. This reduces computational costs by 70\% while improving answer quality. 
Additionally, we will release the Qwen3-32B relevance ranking and utility-based selection annotations for the 100k MS MARCO dataset, providing a high-quality dataset for future research in relevance ranking and utility-based selection. 
For future work, extending utility distillation to tasks beyond QA (e.g., summarization, fact verification) is promising in the future.